# Design of a Timer Queue Supporting Dynamic Update Operations

Zekun Wang
*State Key Discipline Laboratory of Wide Bandgap Semiconductor Technology, School of Microelectronics Xidian University*
Xi'an, China
zekunwang@stu.xidian.edu.cn

Binghao Yue
*State Key Laboratory of Integrated Service Networks Xidian University*
Xi'an, China
23011210684@stu.xidian.edu.cn

Weitao Pan*
*State Key Laboratory of Integrated Service Networks Xidian University*
Xi'an, China
*wtpan@mail.xidian.edu.cn

Jiangyi Shi
*State Key Discipline Laboratory of Wide Bandgap Semiconductor Technology, School of Microelectronics Xidian University*
Xi'an, China
jyshi@mail.xidian.edu.cn

Yue Hao
*State Key Discipline Laboratory of Wide Bandgap Semiconductor Technology, School of Microelectronics Xidian University*
Xi'an, China
yhao@xidian.edu.cn

*Abstract*—Large-scale timers are ubiquitous in network processing, including flow table entry expiration control in software defined network (SDN) switches, MAC address aging in Ethernet bridges, and retransmission timeout management in TCP/IP protocols. Conventional implementations suffer from critical limitations: low timing accuracy due to large-scale timer traversal and high computational overhead for new timer insertion. This paper presents a hybrid-architecture hardware priority queue based on systolic arrays and shift registers for efficient timer queue management. The design uniquely supports five operations: enqueue, dequeue, delete, update, and peek—To the best of our knowledge, it is the first hardware priority queue enabling in-queue priority updates. By leveraging centralized Boolean logic encoding within systolic blocks, the design efficiently generates set/shift control signals while the novel push-first operation ensures FIFO ordering for same-priority timers without additional metadata. Experimental results demonstrate that the design operates at over 400 MHz on FPGAs, achieving a 2.2–2.8× reduction in resource consumption compared to state-of-the-art implementations.

## I. INTRODUCTION

With the exponential growth of network bandwidth and the escalating complexity and programmability of network functions, the hardware acceleration of network functionalities has emerged as a prominent research frontier. Concurrently, large-scale timer systems are being deployed in an ever-expanding array of networking scenarios.

In Software-Defined Networking (SDN), OpenFlow networks decouple the control plane from the data plane, where traffic forwarding relies on flow rules. Due to the upper limit on the number of flow rules stored in the flow table, a timeout mechanism is introduced to remove outdated rules. However, traditional OpenFlow employs fixed timeout values without considering flow characteristics, leading to low efficiency in flow table resource utilization. The work [1] proposes a dynamic approach to adjust the timeout values of individual flow entries, enabling the deletion or update of timer values based on flow behavior.

In the TCP offload engine(TOE)[2] service, TCP protocol processing is integrated into the Network Interface Card (NIC) to offload host CPU workloads. TCP protocol ensures the reliable transmission of the data by acknowledgement and retransmission, which use a variety of timers. The retransmission timer, for instance, plays a critical role in managing lost segments: it sets a timer upon data transmission, and if the corresponding ACK response is not received within the specified timeout interval, the segment is retransmitted— a process known as timeout-based retransmission. This scenario necessitates sophisticated timer management to orchestrate response time coordination and maintain protocol correctness.

The hardware offloading of network functions presents significant challenges for managing large-scale timers in hardware architectures. Upon timer expiration, corresponding actions are triggered based on the application scenario. Conventional management approaches typically instantiate a dedicated timer for each task and, within the timing cycle, iterating through all timers to calculate and determine if expiration has occurred. However, considering the heterogeneity in expiration times, the substantial computational overhead of O(n) traversal, and the inefficiency of performing checks on unexpired tasks, integrating priority queues (PQs) into timer management emerges as a promising solution. This approach orders timer tasks by their expiration times and enables real-time comparison of the queue head element with external clock signals, facilitating efficient timeout detection.

In timer queue management, there are several critical operations: enqueueing new timer tasks, deleting existing tasks, and updating target task timestamps. Maintaining queue scalability while minimizing impact on operating frequency is essential for hardware implementations. Common scalable hardware priority queue (PQ) architectures include binary heaps, systolic arrays, shift registers, and first-in-first-out (FIFO) queues, each with distinct limitations:

- Binary heaps exhibit non-scalable operation cycles proportional to queue depth;
- Systolic arrays incur higher resource overhead than shift registers at identical queue depths;

* Corresponding Author

- Shift register architectures suffer from bus loading bottlenecks as queue size expands;
- FIFO queues face escalating hardware costs with increasing priority levels.

To address these challenges, we propose a hybrid architecture integrating systolic arrays and shift registers for timer PQ implementations. The design aims to support efficient enqueue, delete, and update operations while enabling linear scalability and maintaining high-frequency operation. The main contributions of this paper are as follows:

- A scalable hybrid hardware PQ integrating systolic arrays and shift registers, which supports constant-time operations for enqueue, dequeue, deletion, and dynamic priority update—a capability not found in prior hardware PQ designs.
- The push-first operation, which ensures first-in-first-out FIFO ordering for same-priority tasks without requiring additional flag bits.
- A centralized control architecture with Boolean logic encoding for efficient generation of set and shift signals, reducing combinational logic overhead by leveraging hardware-friendly binary encoding.
- Experimental validation showing that the design achieves 2.2–2.8× lower resource consumption than state-of-the-art implementations at identical queue depths, while operating at clock frequencies exceeding 400 MHz on FPGAs.

## II. Related work

There are two common kinds for managing large-scale timers[4]. Simple Cycle Check Timer (SCC) operates by decrementing each timer in the list by one per iteration; upon reaching zero, the associated event is triggered. This approach suffers from degraded timing accuracy and increased traversal overhead in systems with a large number of timers. In contrast, Multi Level Queue Timer employs a hierarchical structure with queues of varying granularities. When adding a timer, it is imperative to insert the timer into the queue with largest granularity. If the timer is expiring, it should be inserted into the subsequent queue, with this process continuing until the timer is situated in the smallest granularity queue. In each queue, with the exception of the initial timer, each subsequent timer functions solely to record the relative value relative to its preceding timer. However, this scheme consumes substantial computing resources when adding timers, and a large number of adding operations may become the bottleneck of the whole system.

When employing PQ approach, each element need only to save the deadline. An external timer is incremented and compared with the queue head element to determine deadline expiration, eliminating the need for full timer traversal. In many application scenarios, dynamic adjustment of timer values (e.g., increasing or decreasing) is essential. However, current research lacks hardware PQ designs that support in-queue update operations.

The work in [5] proposes a hybrid PQ architecture integrating a hardware-accelerated binary heap. The authors contend that delete and decrease-key operations are infrequently used in practice and their hardware implementation incurs significant resource overhead. As a result, the design supports enqueue and dequeue operations via hardware, while relegating decrease-key and delete operations to software implementation.

The shift-register-based PQ implementation in [6] supports delete operations, but its distributed control architecture—where each Basic Shift Register (BSR) embeds independent logic units—suffers from severe bus loading constraints and substantial hardware costs. The work in [7] proposes a PQ using PIFOs to maintain scheduling order, relying on a shift register that supports only push and pop operations. Similarly, the work in [8] introduces a flow scheduler based on a tree-structured systolic array, the proposal is made to implement a stream scheduler based on a tree systolic array structure, enabling programmable PQ scheduling but limiting operations to push and pop.

A single-instruction-multiple-data PQ is implemented by one-dimensional systolic arrays [9], which supports push, pop, and a specialized replace operation—used exclusively for inserting new elements with the highest priority element replacement.

AnTiQ[10] is a hardware-accelerated PQ specifically designed for timer queue implementations, leveraging systolic blocks to support push, pop, drop, and peek operations via state machine control. Push and pop operations require for different the number of cycles, leading to data collisions that necessitated the integration of additional control logic. Ultimately, a clock frequency of 350 MHz was achieved, but the resource overhead was substantial.

Existing hardware priority queue (PQ) management schemes universally support enqueue and dequeue operations, with some additionally enabling delete operations. However, none of the state-of-the-art designs support dynamic update operations. Some of the architectures are limited by scaling problems, and there are problems of high resource overhead and low operation rate. Given that one-dimensional systolic arrays incur higher resource consumption than shift registers at identical queue depths, and that shift registers face expansion bottlenecks due to bus loading constraints, we propose a hybrid architecture integrating shift registers and systolic arrays—specifically designed to support efficient update operations.

## III. Proposed Architecture

This section begins by outlining the queue operations to be implemented, followed by an overview of the overall architecture. It then proceeds to describe the four proposed hardware operations, explains how to resolve conflicts arising from concurrent operations, and concludes with an elaboration on the coding schemes used to generate set and shift control signals.

### A. Queue Operations

Timer queue elements consist of two fields: ID and DATA, where ID serves to uniquely identify tasks and DATA is utilized for priority sorting. A hardware PQ architecture must support five fundamental operations:

- ENQUEUE: an operation to add a new element to queue.

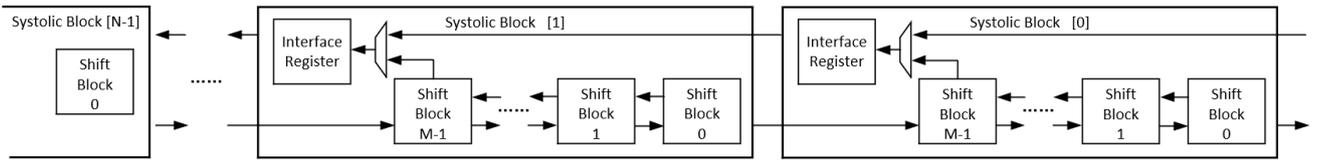

Fig. 1 A hybrid Architecture of systolic arrays and shift blocks

- DEQUEUE: an operation to remove the first element of the queue.
- DELETE: an operation to remove an arbitrary element from the queue.
- UPDATE: an operation to dynamically adjust the DATA field of a target ID within the queue, supporting both incremental and decremental modifications. Namely, adjusting the priority of elements within the queue.
- PEEK: an operation to access the first of the queue without removing it.

Conventional PQs typically implement only enqueue and dequeue operations, which map to push and pop operations in hardware-accelerated PQs. The delete operation, however, requires target ID-based lookup: upon identifying the corresponding element in the queue, it is removed, and all subsequent elements are shifted right.

The update operation serves to dynamically adjust the priority of existing queue elements. For example, the deadline of a task may need to be reduced in the operating system so that it is completed earlier. Alternatively, the response time may be allowed to be larger in the TOE service—both scenarios requiring modification of the priority DATA field. This operation entails locating the target ID and updating its associated DATA, followed by re-sorting the modified element to maintain queue correctness. To the best of our knowledge, no prior hardware priority queue design has supported in-queue updates. This paper presents the first hardware PQ architecture that enables efficient update operations.

### B. Architecture

This paper introduces a hybrid architecture integrating systolic arrays and shift registers for large-scale timer system management. The architecture comprises two core modules: the Systolic Block and the Shift Block. As illustrated in Fig. 1, systolic blocks are serially interconnected to form a systolic array, where each systolic block incorporates M shift blocks. The line segments in Fig. 1 denote the transfer of ID and data fields only. Each element within a shift block comprises the following two fields:

- ID: this field serves as a unique task identifier within the queue, initialized to an all-zero value that denotes an invalid task.
- DATA: this field stores the timing value for each task, which serves as the key for priority-based sorting.

Similar to the function of the temp register in [3], the interface register is used for propagation operations to the next block and can hold either the m-th element or the new push-in element. In this design, the enqueue comparison involves the push-in element, the M elements within the current systolic block, and the first element of the next systolic block—effectively performing a broadcast comparison across M+1 elements.

This is because the update operation may trigger concurrent push and pop operations. When both operations are active simultaneously, push induces leftward shifting while pop causes rightward shifting of elements, leading to repeated manipulation of the next-stage systolic block's head element and exacerbating operational cycle overhead. Thus, direct comparison with the next-block head element reduces cycle overhead. As the first element of each systolic block stage remains in a constant output state, the peek operation requires only external control logic to sample at the clock edge for reading.

The queue is sorted from right to left according to the DATA value, from the highest priority to the lowest priority, with larger values stored to the left of smaller ones. The queue has a maximum capacity of N*M tasks. The hybrid systolic array-shift register architecture offers two key advantages: block-level independence allows only operations propagation between systolic blocks, and this design mitigates bus loading issues in scalable queue implementations.

The internal structure of the shift block is illustrated in Fig. 2. Both [6] and [10] adopt distributed control architectures, where control logic is embedded within each block of the design. The proposed BSR in this paper incorporates three internal comparators, each generating a distinct flag bit. These flag bits are collected within the systolic block and processed through boolean logic operations to achieve centralized control.

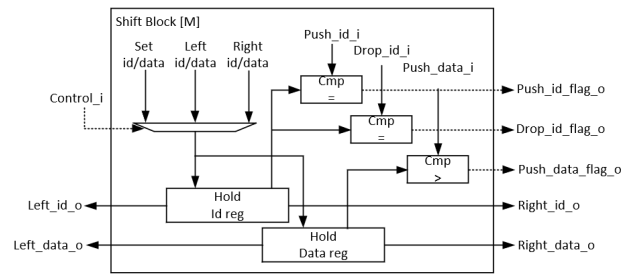

Fig. 2 A single shift block architecture

Equality comparisons between push_id_i, drop_id_i and hold_id are performed to locate elements requiring update or deletion. When push_data_i is less than hold_data, a logic high (1) is generated to denote a suitable insertion point. This less-than comparison mechanism enables systematic search for optimal insertion locations. The hold register stores the current shift block's element and supports modification via set, left-shift, or right-shift operations.

### C. Hardware Operations

To implement the five queue operations described above, specific hardware mechanisms for systolic arrays must be designed. Beyond standard push, pop, and delete operations, this work introduces a novel push-first mechanism for queue hardware—enabling direct insertion of an element into the first shift block of the subsequent systolic block, thereby eliminating the need for comparative operations with other elements. This operation preserves the FIFO order among elements of identical priority. The internal architecture of a single systolic block is illustrated in Fig. 3, and the hardware operations implementation will be elaborated in detail based on this structure.

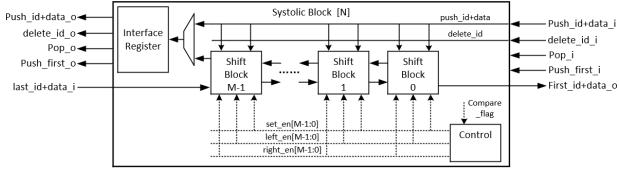

Fig. 3 A single systolic block

*Push Operation*: this operation entails the simultaneous insertion of ID and data into the queue. Since the external initiator has no knowledge of whether the ID exists in the queue, both enqueue and update operations can be implemented via this action. If the ID is already present in the queue, the push action executes an update operation; if not, it performs an enqueue operation. Following the entry of ID and DATA into the queue, both are compared with elements in each shift block through broadcast. Given the need for concurrent ID and data comparison, four distinct scenarios may arise:

- If both the target ID and DATA insertion location of are identified within the current systolic block, no operation is propagated to the next block.

- If only the target ID is found in the current block, but DATA insertion location remains undefined, i.e., the priority of the target ID is updated to become lower, all elements following the target ID in this block are right-shifted. Since the suitable insertion point is not located, the push operation must be forwarded to the next block—meaning push and pop operations are concurrently propagated. When push is active, it can be proven that the pop operation will not propagate to subsequent blocks.

- If the data insertion position is found in the current systolic block and the ID is not found, the element can be inserted into the block, and all subsequent elements are left-shifted. However, since the existence of the same ID in downstream systolic blocks cannot be confirmed, a delete operation is forwarded to remove the target ID element, while the block's last element is pushed to the next block. In other words, both delete and push-first operations are propagated downstream.

- If neither the target ID nor data insertion position is found in the current systolic block, only the push operation propagates to the next block.

*Delete Operation*: This operation searches the queue for the target ID and deletes it. If the target ID is not found, no movement occurs at this block, and a delete operation propagates to the next block. Conversely, upon ID detection, subsequent elements are right-shifted, and a pop operation is forwarded to the next block.

*Pop Operation*: This operation right-shifts all elements within the systolic block and propagates the pop operation to subsequent blocks.

*Push-first Operation*: In the third push scenario, the last element of the current block is propagated into the next block. Since the priority order of the queue must be from highest to lowest, the last element of this block naturally becomes the first element of the next block, obviating the need for repeated comparison in subsequent stages. The next block is updated by left-shifting all elements and setting the first element, a process that guarantees FIFO ordering for tasks with identical priorities but distinct IDs—eliminating the need for auxiliary flag bits[3].

Analysis of operational scenarios reveals that a push operation at the current block may trigger two concurrent operations at the next block. As shown in TABLE I, these two scenarios may induce simultaneous left and right shifts within the same systolic block. Through encoding optimization, both shifts complete in a single cycle, ensuring correct concurrent execution. In TABLE I, ID match denotes the presence of a target ID within the current systolic block. DATA match means finding a suitable insertion position in the current block. This refers to the coexistence of elements with both higher and lower priority than the DATA.

TABLE I OPERATIONS PASSING DUE TO PUSH OPERATION

| Scenarios/ Operations | Push | Delete | Pop | Push-first |
|---|---|---|---|---|
| ID Not Match DATA Not Match | √ | × | × | × |
| ID Not Match DATA Match | × | √ | × | √ |
| ID Match DATA Not Match | √ | × | √ | × |
| ID Match DATA Match | × | × | × | × |

### D. Resolving Conflict

The following two scenarios will trigger concurrent queue operations when a push operation occurs:

- When the ID match without DATA match, push and pop operations may execute concurrently in subsequent blocks. Since the ID matches has already occurred, the push operation can only left-shift the element, while the pop operation can only right-shift the element. These shifts neutralize each other, streamlining conflict resolution control logic.

- When the DATA match without the ID match, both the delete and push-first operations may occur at the later blocks.

The design specifies that the four operations above require four cycles for completion, as illustrated in Fig. 4. Each operation in systolic block N is synchronized with the push operation in cycle count, which decomposes into four phases: enable, compare, set-and-shift, and finish. The push operation outputs delete, pop, or push-first operations to the next block during the fourth cycle.

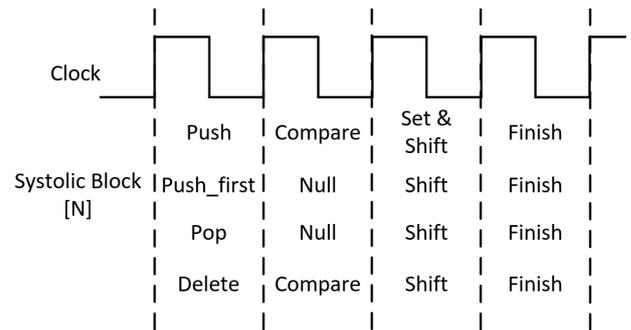

Fig. 4 Timing Graph of every operation

The pop operation requires no a comparison process and, in principle, completes in fewer cycles. However, during propagation through subsequent systolic blocks, concurrent pop and push operations may occur. The push operation necessitates that all elements in the current systolic block and the next block's first element remain in stable positions during the comparison phase. If the next block's pop operation is in the shifting phase or incomplete shift state, this can cause push operation failure in the current block. Consequently, the design incorporates a null operation during the pop operation to align its cycle count with the four-cycle push operation.

During the comparison phase of the push operation, in addition to performing a broadcast comparison within the current block, a comparison must also be made with the first element of the next systolic block to determine whether the current inserted DATA should be placed in the next systolic block or the last shift block of the current block. This prevents scenarios where, despite failing to find a valid insertion position for DATA at the current block, an active pop operation at this block misdirects the element—destined for the current-systolic block's last shift block—into the first shift block of the systolic block, as illustrated in Fig. 5.

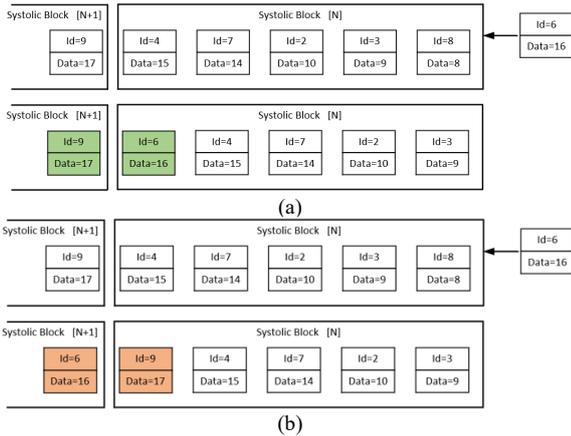

Fig. 5 Push and pop happening at the same time: (a) comparing the first element in the next block and (b) not comparing

As illustrated in Fig5(b), concurrent push and pop operations at the current block may lead to priority ordering errors if not compared with the next block's first element. The correct ordering is illustrated in Fig. 5(a). The red squares in Fig. 5(b) exhibit priority ordering errors, whereas the green squares in Fig. 5(a) are correctly ordered.

When both delete and push-first operations are active, it is ensured that all elements subsequent to the ID match remain unmodified, while all elements preceding the ID match are left-shifted to overwrite the ID match 's position.

The architecture proposed in this paper mandates a minimum four-cycle interval between adjacent operations, as depicted in Fig. 6. When the finish phase of the (N-1)th systolic block issues a push operation to the next block, and the finish phase of the Nth systolic block stabilizes its first element, the (N-1)th block can correctly execute the compare process in the subsequent push operation. Thus, a minimum interval of four cycles is mandated—meaning each operation requires five cycles to complete. The systolic block completes the queue operations recursively, and all operation completion times are constant.

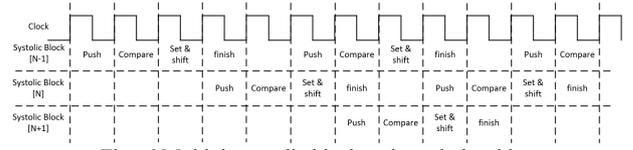

Fig. 6 Multiple systolic block action relationships

*E. Solving for shift and set signals*

The key to the control logic inside the systolic block is how to determine the correct element insertion position and shift signal. We employ centralized control over all shift blocks in a systolic block, incorporating boolean logic operations to resolve shift and placement tasks. This approach streamlines implementation logic while reducing resource overhead.

During the implementation of the delete or push operation, if the ID match is successful, the result will yield a one-hot code, whereas a successful data comparison results in a binary sequence of consecutive 0s and 1s. Set and shift control signals are derived from boolean operations that resolve the positions of binary digits in these sequences. When implemented via priority coding, this approach incurs significant resource overhead. A left or right shift of other elements is possible when the push operation successfully matches an ID and data within a systolic block. These scenarios will be illustrated with examples respectively.

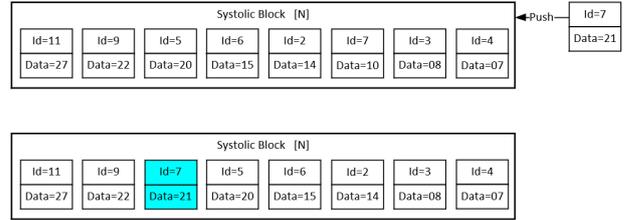

Fig. 7 Right shift due to push

As shown in Fig. 7, input an element with ID=7 and DATA=21. The matching result of ID is *id_flag*, and the result of the data comparison is *data_flag*. The set signal is derived from the data comparison result: since the insertion position lies to the left of the ID matching position, a right-shift operation is triggered. The data comparison result is right-shifted, and the highest bit is set to 1'b1 to generate *data_flag_lp*:

$$data\_flag\_lp = \{1'b1, data\_flag[M-1:1]\} \quad (1)$$

Subtracting 1'b1 and then inverting the result generates the set signal *set_en*:

$$set\_en = \sim(data\_flag\_lp - 1'b1) \quad (2)$$

The computation of the right-shift enable signal *right_en* integrates the ID match result and data comparison result. This is achieved by first subtracting 1'b1 from the ID match result, then performing an XNOR operation with *data_flag_lp* to generate *right_en*:

$$right\_en = data\_flag\_lp \odot (id\_flag - 1'b1) \quad (3)$$

Thus, after pushing an element with ID=7 and DATA=21, the set signal is: 8'b0010_0000, and the shift signal is 8'b0001_1100. The blue squares in Figure 7 represent the updated positions of the elements.

When pushing an element with ID=9 and DATA=9, as illustrated in Fig. 8, the ID match position is situated to the left of the DATA match position. Consequently, some elements must undergo left-shift operations. The set signal is derived by subtracting 1'b1 from the data comparison result and then inverting it:

$$set\_en = \sim(data\_flag - 1'b1) \quad (4)$$

The left-shift enable signal *left_en* is generated by first subtracting 1'b1 from the ID match result, performing an XNOR operation with the data comparison result, and appending 1'b0 to the right of the result.

$$left\_en = \{(data\_flag \odot (id\_flag - 1'b1)), 1'b0\} \quad (5)$$

Therefore, after enqueuing an element ID = 9 and DATA=9, the set signal is 8'b0000_0100, and the shift signal is 8'b0111_1000. The blue squares in Figure 8 represent the updated positions of the elements.

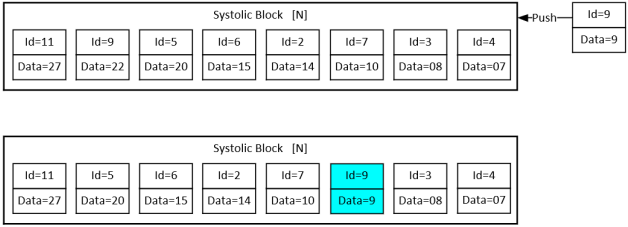

Fig. 8 Left shift due to push

The set and shift signals of push-first, delete, and pop are derived using a similar approach.

## IV. EXPERIMENTAL RESULTS

### A. Prototype Implementation

The proposed hardware PQ was implemented in Verilog, and its verification was conducted using System Verilog (SV). In the SV platform, test incentive IDs and DATA sequences with more than twice the number of queue depths are randomly generated. The golden model, also implemented in SV, was employed to compare the results with the queue results of sorting completion in the DUT.

FPGA synthesis and implementation of the design was performed with Xilinx Vivado 2024.2, with the Xilinx VCU 118 specified as the target platform. The design is implemented using asynchronous reset, and the reconfigurable parameters include : ID width, DATA width, the number of systolic blocks N, and the number of shift blocks in each systolic block M. The ID width, determined by the queue depth, is configured as $clog2 (queue_depth) = $clog2 (N*M).

### B. Cost And Performance

In Experiment I, M was kept constant while N is varied to investigate whether the number of logic levels scales with N and to validate the systolic array characteristic.

For a data width of 16 bits, each systolic block incorporates 16 shift blocks with M fixed at 8. The number of systolic blocks (N) is systematically varied, and the maximum logic level for each architectural configuration is recorded.

As shown in Table II the maximum number logic level remains consistent across different queue depths, confirming that the design adheres to the characteristics of systolic arrays. In terms of resource utilization, the consumption of LUTs and FFs increases linearly with the queue depth. The clock frequency varies across different queue depths due to the wider ID width required for deeper queues, which complicates the combinational logic for comparison. At a queue depth of 256, the design achieves a clock frequency of 469 MHz. Compared with the design in [10], this implementation demonstrates both higher performance and lower resource overhead.

TABLE II IMPLEMENTATION RESULTS OF COMPETING ARCHITECTURES UNDER FIXED DATA WIDTH AND M

| Queue Depth | ID width | DATA width | N | Logic Level | LUTs | FFs | Fmax (MHz) |
|---|---|---|---|---|---|---|---|
| 4096 | 11 | 16 | 512 | 5 | 279544 | 207855 | 337 |
| 3072 | 11 | 16 | 384 | 5 | 209637 | 155878 | 333 |
| 2048 | 10 | 16 | 256 | 5 | 139834 | 100325 | 381 |
| 1536 | 10 | 16 | 192 | 5 | 104959 | 75237 | 382 |
| 1024 | 9 | 16 | 128 | 5 | 66633 | 48359 | 416 |
| 640 | 9 | 16 | 80 | 5 | 41653 | 30215 | 439 |
| 256 | 8 | 16 | 32 | 5 | 16033 | 11179 | 469 |

In Experiment II, the ID width was fixed at 9 bits and N was set to 32. By systematically varying the queue depth and data width, we aim to characterize the resource consumption of proposed PQ architectures.

TABLE III IMPLEMENTATION RESULTS OF COMPETING ARCHITECTURES UNDER FIXED ID WIDTH AND N

| Queue Depth | ID width | DATA width | M | Logic Level | LUTs (%) | FFs (%) | Fmax (MHz) |
|---|---|---|---|---|---|---|---|
| 512 | 9 | 16 | 16 | 5 | 31442 | 19561 | 431 |
| 320 | 9 | 16 | 10 | 5 | 20554 | 13609 | 433 |
| 256 | 9 | 16 | 8 | 5 | 16600 | 11625 | 431 |
| 128 | 9 | 16 | 4 | 5 | 8955 | 7657 | 478 |
| 512 | 9 | 64 | 16 | 7 | 70740 | 48745 | 404 |
| 320 | 9 | 64 | 10 | 7 | 46612 | 33577 | 399 |
| 256 | 9 | 64 | 8 | 7 | 38332 | 28521 | 423 |
| 128 | 9 | 64 | 4 | 8 | 21923 | 18409 | 408 |

As shown in Table III, when the ID width and DATA width are fixed, varying the number of shift blocks within a single systolic block reveals a more pronounced decrease in LUT count with decreasing queue depth. This indicates that the majority of combinational logic overhead in centralized control resides within the systolic block. Notably, quadrupling the data width results in a total area increase of less than 2.5 times, demonstrating that the centralized control scheme outperforms the distributed state machine control [10]. However, increasing the data width leads to a decline in operating frequency, as encoding calculations significantly impact the maximum fan-out.

A comparative analysis between the proposed architecture and the design in [10] was performed with a fixed data width of 16 bits (where the ID width equals the data width). Timing closure failed when the clock period was constrained to 2.5 ns, necessitating an adjustment to 2.8 ns. As tabulated in TABLE IV, the proposed design demonstrates around 20% higher frequency with an average reduction of 37% in LUTs usage and 36% in FFs usage, translating to a 2.2–2.8× area reduction.

TABLE IV IMPLEMENTATION OF ANTIQ IN [10]

| Queue Depth | ID width | DATA width | LUTs | FFs | Fmax (MHz) |
|---|---|---|---|---|---|
| 512 | 16 | 16 | 82232 | 60894 | 364 |
| 320 | 16 | 16 | 56614 | 38046 | 364 |
| 256 | 16 | 16 | 45323 | 30403 | 363 |
| 128 | 16 | 16 | 22631 | 15198 | 365 |

## V. CONCLUSION

This paper presents a scalable hardware PQ that employs a hybrid architecture integrating systolic arrays and shift registers. By leveraging the bus-load-free nature and constant operation latency of systolic arrays, along with the resource efficiency of shift registers, the proposed design supports not only standard queue operations (enqueue, dequeue, delete, peek) but also priority updates for existing tasks. In terms of hardware operations, we introduce the push-first operation, which ensures FIFO order for tasks with identical priorities without additional flag bits. Resource overhead is reduced through centralized control and Boolean logic encoding within the systolic blocks to generate set and shift signals. With reconfigurable parameters, the design achieves clock frequencies exceeding 400 MHz for large-scale queues on FPGAs. In the future, we aim to reduce resource overhead and minimize the cycle interval between operation initiations without performance degradation by introducing modules such as RAM macro cells. The design will even be integrated into a complete network processor to evaluate its capabilities and performance in real high-speed networking systems.

## ACKNOWLEDGMENT

This work is supported by the National Key R&D Program of China (2023YFB4405100, 2023YFB4405102).